%% file: opensource_hw_date.tex
\documentclass[journal,10pt]{IEEEtran}

\pdfoutput=1
\ifCLASSOPTIONcompsoc
  \usepackage[nocompress]{cite}
\else
  \usepackage{cite}
\fi
\ifCLASSINFOpdf
\else
\fi
\usepackage{amssymb}
\usepackage[cmex10]{amsmath}
\usepackage{algorithmicx}

\usepackage{url}
\usepackage{amssymb}
\usepackage{booktabs}
\usepackage{url}
\usepackage[normalem]{ulem}
\usepackage{color}
\usepackage[dvipsnames]{xcolor}
\usepackage{adjustbox}
\usepackage{framed}

\usepackage{graphicx}
\usepackage{enumitem}
\usepackage{epstopdf}
\usepackage{booktabs}
\usepackage{multirow}
\usepackage{color}
\usepackage{algorithm}
\usepackage{algpseudocode}

\begin{document}
\bstctlcite{IEEEexample:BSTcontrol}

%
% paper title
% Titles are generally capitalized except for words such as a, an, and, as,
% at, but, by, for, in, nor, of, on, or, the, to and up, which are usually
% not capitalized unless they are the first or last word of the title.
% Linebreaks \\ can be used within to get better formatting as desired.
% Do not put math or special symbols in the title.
\title{\textit{OpenHealth}: Open Source Platform for Wearable Health Monitoring}

\author{Ganapati Bhat, Ranadeep Deb, Umit Y. Ogras \\
	Email: \{gmbhat, rdeb2, umit\}@asu.edu \\
	School of Electrical, Computer and Energy Engineering, Arizona State 
	University, Tempe, AZ
	\thanks{ This article has been accepted for publication in a future of IEEE 
	Design \& Test of Computers}
\thanks{\copyright 20XX IEEE.  Personal use of this material is permitted.  
	Permission from IEEE must be obtained for all other uses, in any current or 
	future media, including reprinting/republishing this material for 
	advertising 
	or promotional purposes, creating new collective works, for resale or 
	redistribution to servers or lists, or reuse of any copyrighted component 
	of 
	this work in other works.}
}

\IEEEtitleabstractindextext{%
\begin{abstract}

Movement disorders are becoming one of the leading causes of functional disability 
due to aging populations and extended life expectancy. 
Wearable health monitoring is emerging as an effective way to augment clinical care for movement disorders. 
However, wearable devices face a number 
of adaptation and technical challenges that hinder their widespread adoption. 
To address these challenges, we introduce \textit{OpenHealth}, 
an open source platform for wearable health monitoring.  
\textit{OpenHealth} aims to design a standard set of hardware/software and wearable devices 
that can enable autonomous collection of clinically relevant data. 
The \textit{OpenHealth} platform includes a wearable device, 
standard software interfaces and reference implementations
of human activity and gesture recognition applications.
\end{abstract}

% Note that keywords are not normally used for peerreview papers.
}

% make the title area
\maketitle

% To allow for easy dual compilation without having to reenter the
% abstract/keywords data, the \IEEEtitleabstractindextext text will
% not be used in maketitle, but will appear (i.e., to be "transported")
% here as \IEEEdisplaynontitleabstractindextext when the compsoc
% or transmag modes are not selected <OR> if conference mode is selected
% - because all conference papers position the abstract like regular
% papers do.
\IEEEdisplaynontitleabstractindextext
% \IEEEdisplaynontitleabstractindextext has no effect when using
% compsoc or transmag under a non-conference mode.

% For peer review papers, you can put extra information on the cover
% page as needed:
% \ifCLASSOPTIONpeerreview
% \begin{center} \bfseries EDICS Category: 3-BBND \end{center}
% \fi
%
% For peerreview papers, this IEEEtran command inserts a page break and
% creates the second title. It will be ignored for other modes.
\IEEEpeerreviewmaketitle

%\IEEEraisesectionheading{\section{Introduction}\label{sec:introduction}}
% Computer Society journal (but not conference!) papers do something unusual
% with the very first section heading (almost always called "Introduction").
% They place it ABOVE the main text! IEEEtran.cls does not automatically do
% this for you, but you can achieve this effect with the provided
% \IEEEraisesectionheading{} command. Note the need to keep any \label that
% is to refer to the section immediately after \section in the above as
% \IEEEraisesectionheading puts \section within a raised box.

\input{files/introduction.tex}
\input{files/wearable_barriers.tex}
\input{files/openhealth_overview.tex}

\input{files/opensource_hw.tex}
\input{files/software_stack.tex}
\input{files/applications.tex}
\input{files/conclusion_future_research.tex}

\input{opensource_hw_date.bbl}

%\bibliographystyle{IEEEtranS}
%{\footnotesize{\bibliography{references/embedded_refs,references/flexible,references/health_refs}}}

%\vspace{-1cm}
\begin{IEEEbiographynophoto}
		{Ganapati Bhat} received his B.Tech degree in Electronics and 
		Communication from Indian School of Mines, Dhanbad, India in 2012. He is currently a Ph.D. student in Computer Engineering at the school of Electrical, Computer and Energy engineering, Arizona State University. He is a student member of the IEEE and ACM.
\end{IEEEbiographynophoto}
\begin{IEEEbiographynophoto}{Ranadeep Deb} received his B.Tech in Electronics and Communication engineering from WBUT, India. He is currently pursuing MS in Computer Engineering at Arizona State University with a specialization in VLSI Design. He is working on applications on wearable sensors in assessment and prediction of human activities and monitoring of biomarkers in Parkinson's disease patients.
\end{IEEEbiographynophoto}
\begin{IEEEbiographynophoto}{Umit Y. Ogras} received his Ph.D. degree in Electrical 
and Computer Engineering from Carnegie Mellon University, Pittsburgh, PA, in 2007. From 
2008 to 2013, he worked as a Research Scientist at the Strategic CAD 
Laboratories, 
Intel Corporation. He is currently an Assistant Professor at the School of 
Electrical, Computer and Energy Engineering.
\end{IEEEbiographynophoto}

\end{document}

%% file: files/introduction.tex
\section{Introduction}

Coupled with an aging population and extended life expectancy, movement disorders 
are becoming one of the leading causes of functional disability~\cite{dorsey2016moving}.
For example, Parkinson's disease~(PD), essential tremor~(ET), epilepsy, and stroke 
affect more than 70 million people worldwide~\cite{daneault2018could}. 
%It is estimated that as many as 8.7 million people will suffer 
% from PD in the most populous nations by 2030~\cite{dorsey2016moving}.
%
%Wearable internet-of-things (IoT) devices and smart health technology 
%offer a great potential to empower this population to perform daily functions, 
%while improving the quality of life of healthy individuals~\cite{espay2016technology,yin2018smart}.
%
%
Diagnosis and treatment of movement disorders currently rely on 
tests and observations made by specialists in a medical facility, who prescribe medication and therapy based on these observations~\cite{maetzler2016clinical}. 
However, clinical visits, which are typically weeks apart, 
capture only a snapshot of the symptoms~\cite{estrin2010open}. 
This introduces complications in therapy decisions, 
since symptoms vary over time, and patients' recall accuracy is not reliable~\cite{ozanne2018wearables}.
%For example, PD has a number of motor and nonmotor symptoms 
%that differ in each patient and have varying degrees of 
%progression in each patient~\cite{maetzler2016clinical,heldman2017telehealth}.
%Similarly, recall accuracy can be highly variable in clinical visits. 
%This leads to challenges in monitoring of patients with epilepsy due to the 
%unpredictable nature of seizures and difficulty in remembering the occurrence 
%of such seizures~\cite{ozanne2018wearables}.
Moreover, access to highly trained specialists can be challenging in many parts of the world. 
%For example, China has only about 50 movement disorder specialists for over two million patients with PD~\cite{dorsey2016moving}.
%

Wearable sensors and mobile health applications are emerging as attractive solutions to augment clinical
treatment and enable telepathic diagnostics~\cite{espay2016technology,estrin2010open,ozanne2018wearables,daneault2018could}.
Wearable technology allows for continuous monitoring of user movement 
in a free-living home environment. 
This capability helps in capturing the progression of symptoms that change over time. 
Furthermore, it enables evaluating the prescribed therapy 
on an individual basis~\cite{matias2017perspective,maetzler2016clinical,heldman2017telehealth}.
%
%In addition to symptom monitoring, wearables can track how patients are doing 
%in their daily life activities 
%
Similarly, wearable sensors and smartphones have shown promising results in 
the diagnosis of ET~\cite{daneault2018could} and 
detecting seizures in epilepsy~\cite{ozanne2018wearables}. 
Studies have also shown that both patients and health professionals (HPs) value the interactive 
information available from wearable monitoring. %~\cite{ozanne2018wearables}.
Hence, wearable sensors coupled with telepathic diagnostics 
can greatly improve health care~\cite{heldman2017telehealth,ozanne2018wearables}. 
%In summary, wearable technology has already shown promising results 
%in the diagnosis and treatment of movement disorders.

Despite the promising results demonstrated so far, widespread adoption of wearable technology is hindered by both technology and adaptation challenges. 
The International Parkinson and Movement Disorders Society Task Force on Technology 
identifies the major challenges as non-compatible platforms, 
clinical relevance of the ``big data'' acquired by sensors, 
and wide-spread/long-term deployment of new technologies~\cite{espay2016technology}.  
According to the task force, open source projects can help in addressing these 
challenges by providing a common platform, with standardized hardware~(HW) and 
software~(SW) tools, driven by burning clinical needs.
Several open source solutions for medical devices have been 
proposed recently. The work in~\cite{niezen2016open} surveys open source 
devices for infusion pumps, brain-computer interfaces, CT scanners, and 
physiological monitoring. Among these, the 
e-Health\footnote{https://www.cooking-hacks.com/documentation/tutorials/ehealth-biometric-sensor-platform-arduino-raspberry-pi-medical}
sensor platform is the most relevant for movement disorders, as it provides sensors for monitoring motion. 
However, the e-Health sensor platform comes in a large form factor, making it 
unsuitable for long-term wearable use.

The goal of this work is to discuss the major barriers to widespread deployment of wearable health technology 
and present the \textit{OpenHealth} framework as a potential solution.
\textit{OpenHealth} is an open source HW/SW platform for wearable health monitoring.
\textit{Our vision is to bridge the gap between isolated research activities, health professionals, and technology developers 
by facilitating research using a common platform, standards, and data sets.}
Our open source release\footnote{
\url{https://sites.google.com/view/openhealth-wearable-health/home}} includes all the HW and SW files of the OpenHealth platform, 
which includes energy harvesting circuitry, 
a modular sensor hub, 
processing hardware, a wireless modem, software drivers, and application-programming interfaces (API).
Our initial application area is activity monitoring for movement disorder patients.
Future versions of our open source platform will include applications such as fall detection and seizure detection in epilepsy. 
We also provide 
reference implementations and data sets for human activity and gesture recognition applications.
This feature aims to enable researchers who focus on applications to use 
\textit{OpenHealth} without dealing with hardware details. 
All the hardware design files, software repository, and data sets are released under the GNU 
General Public License. 
%One of the major challenges is that most of currently available solutions are non compatible 
%with each other, making it difficult combine the data generated by different devices. 
%This leads to further difficulties in using the data for making therapy decisions~\cite{espay2016technology}.
%In order to bridge this gap, we propose 

The rest of this paper is organized as follows. 
Section~\ref{sec:barriers} overviews the challenges faced by wearable technologies and our solution strategies. 
Our vision for the operation of \textit{OpenHealth} and implementation details of our current release 
are discussed in Section~\ref{sec:openhealth_overview} and Section~\ref{sec:hardware}, respectively. 
Finally, Section~\ref{sec:applications} presents two reference applications,
and Section~\ref{sec:conclusion} summarizes our future directions.

%% file: files/wearable_barriers.tex
\section{Wearable Health: Challenges and Solutions} \label{sec:barriers}

\begin{figure}[b]
	\centering
	\includegraphics[width=1.0\linewidth]{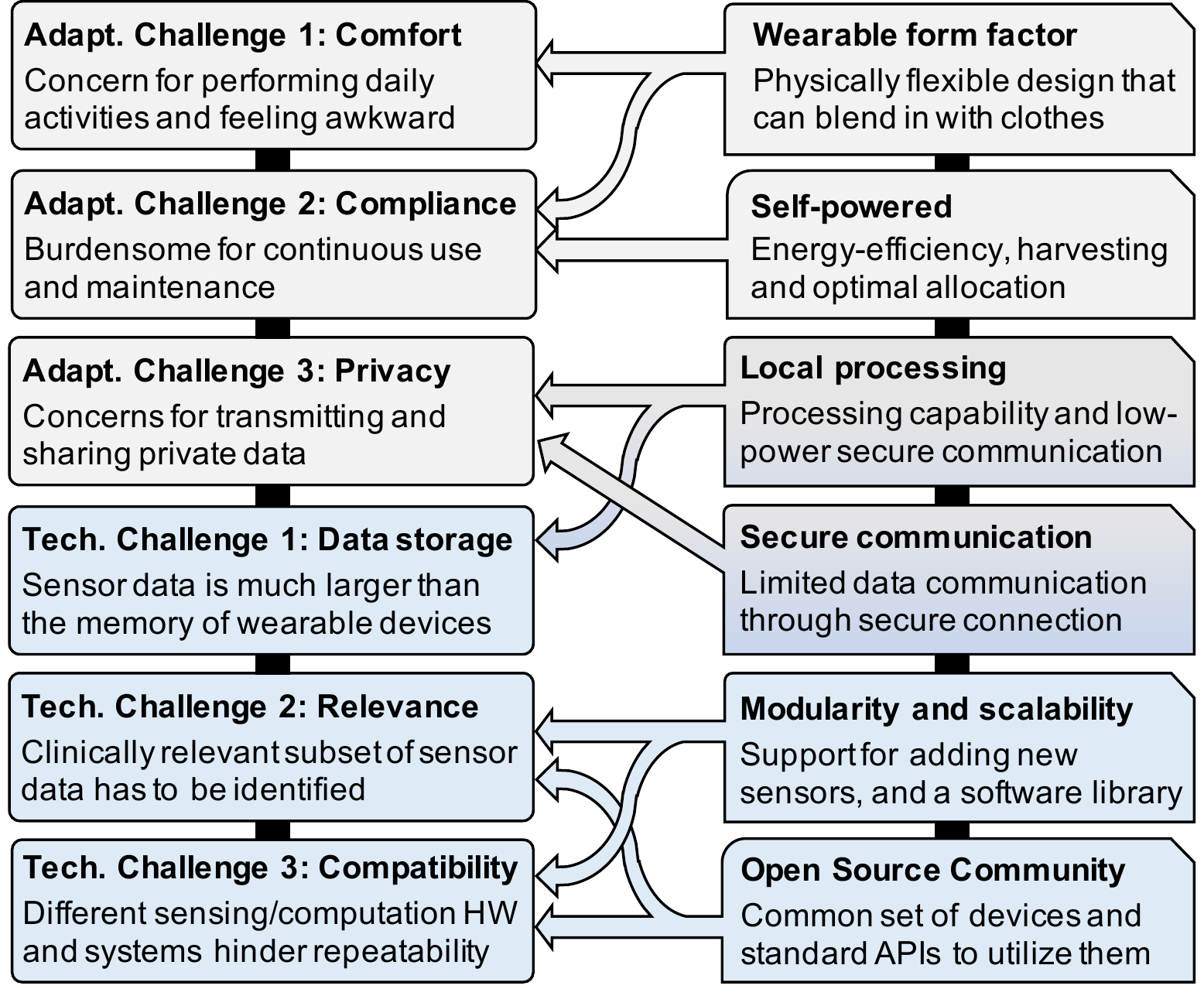}
	\caption{Challenges of wearable health platforms and proposed strategies}
	\label{fig:challenges}
\end{figure}

%Several major challenges need to be addressed before wearable technology can be widely
%adopted~\cite{maetzler2016clinical,daneault2018could,ozanne2018wearables}. 
%We classify them broadly as adaption and technology challenges, 
%as summarized in Figure~\ref{fig:challenges}. 

We classify the barriers to widespread adaptation of wearable health 
technologies 
as adaptation and technical challenges, as detailed in 
Figure~\ref{fig:challenges}.
The adaptation challenges include social and user-specific barriers 
that prevent widespread deployment. 
In contrast, the technical challenges include 
barriers faced by the designers of the wearable platforms. 
%This section overviews these challenges and points at the strategies 
%adopted by the \textit{OpenHealth} platform without delving into details. 
%The operation of the open source vision, 
%and concrete details of the current release are presented in the following two sections. 

\noindent \textbf{Adaptation Challenge 1 - Comfort:}  
A substantial number of patients who participated in previous studies 
have reported feeling self-conscious when using wearable devices. 
They anticipate that the wearable device might be 
a probable cause of stigmatization~\cite{johansson2018wearable,ozanne2018wearables}.
The users also expressed feeling embarrassment and awkwardness when wearing the sensor in public. 
It was stressful to even wear the sensor for some patients~\cite{johansson2018wearable,ozanne2018wearables}. 
To address this challenge, we propose devices based on flexible hybrid electronics, 
as illustrated in Figure~\ref{fig:prototype_photo}a and detailed in 
Section~\ref{sec:base_hw}.
%Ganapati: Pls make sure that each paragraph has this type of a link}. 
%
This approach enables physically flexible or stretchable devices that can blend in with clothes, 
such as a knee sleeve~\cite{bhat2018online}.

\noindent \textbf{Adaptation Challenge 2 - Compliance:}  
Participants in prior studies reported difficulty in using and charging the device regularly.  
It is difficult for a patient to charge the device, 
since it may involve taking the device off and wearing it on again~\cite{ozanne2018wearables}. 
Larger and bulky batteries help in increasing the lifetime of the device, but they also make the device uncomfortable to wear. 
%Moreover, patients with limited motor skills experience difficulties in keeping 
%the device in the correct position~\cite{johansson2018wearable}.
%
%Furthermore, the participants worried that they might damage the wearable while 
%performing activities of daily life (ADL) like washing dishes or showering~\cite{johansson2018wearable}.
%Another barrier for wider adaption of wearables is patient compliance to home monitoring. 
Studies have also shown that many patients find wearable devices uncomfortable or burdensome 
and stop using them after some time~\cite{daneault2018could}. 
%l{sec:base_hw}
Thus, the device should be able to operate autonomously with minimum human intervention. 
%To improve sustainability,
Therefore,
\textit{OpenHealth} includes energy harvesting and dynamic energy 
allocation, which can eliminate battery charging requirements~\cite{bhat2017near}. 
Furthermore, \textit{OpenHealth} platform operates autonomously to facilitate 
use without human intervention. 
More specifically, it automatically turns on, manages the power states 
and communicates with a host device, such as a smartphone, when it senses motion,  
as described in Section~\ref{sec:base_hw}. 
Physically flexible form factor also promotes compliance, as illustrated in Figure~\ref{fig:challenges}. 
%wearable design and maintenance-free operation address this challenge. 

\noindent \textbf{Adaptation Challenge 3 - Privacy:}  
Prior studies have shown that data privacy and security are among the primary concerns 
about using wearable devices for health monitoring~\cite{ozanne2018wearables}. 
Raw sensor data could be transferred to the cloud for identification of technology-based objective measures (TOMs). 
This can cause security pitfalls as the sensor data contains sensitive information about the patient's health. 
The first solution to this concern is processing the user-specific data locally to the maximum extent possible. 
For example, motion data from PD patients is processed locally to extract 
clinically relevant information. 
Then, \textit{only the processed data} is transmitted through a secure channel 
\textit{only to the health professional in charge}.
%In order to address this challenge, we process the sensor data locally on the 
%wearable device and transmit \emph{only} the result to a host device using an
%encrypted BLE connection.

\noindent \textbf{Technical Challenge 1 - Data Storage: }  
Wearable sensors can collect a large amount of data. 
For instance, a 3-axis accelerometer alone can collect more than 5MB of data in one hour, 
while the  local storage capacity of wearable devices is in the order of few megabytes. 
Hence, long-term storage of raw data is not sustainable. 
Since transmitting the raw data would quickly deplete the battery, it is not a viable option either. 
Therefore, the proposed solution provides local processing capability, 
as well as a library of signal processing and machine learning algorithms, as 
detailed in Sections~\ref{sec:base_hw} and~\ref{sec:software_stack}. 
This strategy also benefits the privacy challenge.

%With the rise of big data and cloud computing, increasing amount of data 
%is being sent to the cloud for processing~\cite{farahani2018towards}. 
%As a result, there are concerns about  the security and privacy of the data. 
%In the context of health monitoring, raw  sensor data from patients is transferred 
%to the cloud for identification of TOMs. The results are then communicated back to the wearable device. 

\begin{figure*}[t]
	\centering
	\includegraphics[width=0.90\linewidth]{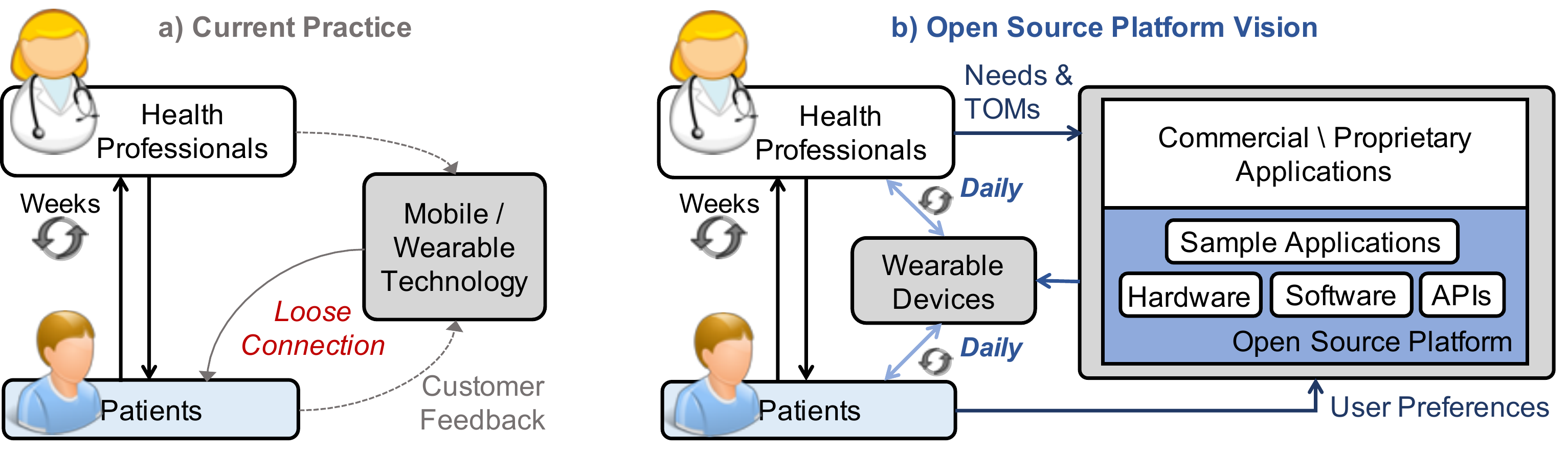}
	\caption{(a) Current practice of healthcare for movement disorders (b) 
		Envisioned solution with the use of open source wearable platform}
	\label{fig:health_scenario}
\end{figure*}

\noindent \textbf{Technical Challenge 2 - Relevance of Sensor Data: }
Large amounts of sensor data do not necessarily mean that all the data are clinically relevant~\cite{espay2016technology}. 
% 
% Ganapati: How are unstructured/crowd-sourced channels relevant to our paper? 
%For example, large amount of data collected from unstructured and crowd-sourced channels 
%does not always translate to relevant clinical metrics~\cite{espay2016technology}.
In fact, a high volume of data can dilute its direct applicability~\cite{maetzler2016clinical}. 
Hence, sensors and algorithms should effectively extract relevant information for individualized patient treatment~\cite{maetzler2016clinical,daneault2018could}.
We address this challenge through two mechanisms. 
First, the proposed platform features a modular sensor hub that allows 
adding new sensors for specific use-cases. 
%Similarly, most wearable solutions for monitoring of movement disorders require 
%some signal processing and machine learning algorithms~\cite{matias2017perspective}. 
Furthermore, the proposed platform provides hardware support and built-in functions, 
such as a variety of filtering and bio-marker generation algorithms. 
Second, the health professionals are principal members of the open source community. 
They communicate the clinical needs and TOMs with the developers, 
as shown in Figure~\ref{fig:health_scenario} and detailed in Section~\ref{sec:openhealth_overview}. 
%Consequently, hardware and software required for extracting clinically relevant data is integrated to the open source platform. 
%For long-term monitoring, often the 
%participants are trusted with the supervised learning and it is expected that 
%they will keep a detailed record of their symptoms, which can be misleading and suffer from recollection bias~\cite{lopez2016inertial}.
%Similarly, wearable devices typically have limited storage capabilities, which requires transmission of data to the cloud. 
%This further raises questions for data privacy and security~\cite{ozanne2018wearables}.
%Therefore, future wearable devices should have efficient algorithms that can 
%provide clinically relevant bio-markers and have strong protocols for data security.

\noindent \textbf{Technical Challenge 3 - Compatibility: }
The International Parkinson and Movement Disorders Society Task Force on 
Technology emphasizes that the majority of technology development efforts
operates within its own ``islands of expertise'', with limited compatibility 
among the systems~\cite{espay2016technology}. 
Since devices from distinct manufacturers may give non-compatible results, 
it is difficult to integrate data provided by different devices. 
In order to bridge this gap, we propose an open source design methodology where 
the wearable devices are derived from a common base platform, 
as illustrated in Figure~\ref{fig:health_scenario}(b) and detailed in Section~\ref{sec:hardware}. 
The compatibility of the proposed solution is improved by using the same underlying hardware,
preprocessing software and standard interfaces. 
This also facilitates comparing results from different research groups. 
Finally, open source solutions can constitute the foundation for commercial products, 
which add new proprietary intellectual property (IP) on top of the commonly used solutions. 

\vspace{1mm}
\noindent \textbf{Reliability and robustness:} In addition to addressing the adaptation and technical challenges, we must ensure that wearable devices are reliable and robust. Reliable and robust design of the device ensures that the device remains operational when subjected to stress during normal use by patients. Common causes of stress for wearable devices include bending, rolling, and folding by patients as a result of their activities~\cite{gupta2017flexibility}. 
We simulate different bending patterns in a finite element simulator~(COMSOL) 
to test the reliability of the device before the manufacturing process. 

The device must also be able to continuously sample the sensors, generate the features, and notify a caregiver in case emergencies. This is especially important for 
life-threatening of emergencies such as seizures in patients with epilepsy. In our proposed solution, we aim at continuous energy-neutral operation by incorporating energy harvesting and a backup battery in the \textit{OpenHealth} platform~\cite{bhat2017near}. 
Furthermore, we perform on-device processing of the sensor data such that the latency of transferring the sensor data to a host device can be avoided.

%% file: files/openhealth_overview.tex
\section{OpenHealth Vision} \label{sec:openhealth_overview}
% Paragraph 1 goal: Describe the current healthcare scenario
%In today's clinical environment, 
Patients with movement disorders primarily interact with health professionals 
during office visits, which are typically weeks apart. 
Recently, smartphone apps and fitness trackers have been employed 
to monitor patients' symptoms during their daily life~\cite{estrin2010open}, 
as depicted in Figure~\ref{fig:health_scenario}a. 
While this is a useful starting point, these devices are not designed
to take into consideration the patient needs.
Instead, the patients and health professionals provide feedback after using the device. 
Hence, they do not address the social and technical challenges, 
such as comfort, compliance, and clinical relevance, as discussed in 
Section~\ref{sec:barriers}.
Consequently, the connection between the patients' needs and device capabilities is loose.
%Instead, there is only a loose connection between the developers of the devices and their users. 

\textit{OpenHealth} aims to provide a comprehensive framework that enables a tighter and systematic interaction 
between all the stakeholders in wearable health monitoring, as illustrated in Figure~\ref{fig:health_scenario}b. 
Open source communities can bridge isolated research efforts, health professionals, and HW/SW developers
to address the social and technical challenges~\cite{espay2016technology}. 
% to work together in developing devices that meet the 
%needs of health professionals and preferences of the patients that use them.
% OpenHealth open source platform. 
In this framework, health professionals provide needs and clinically relevant TOMs to the developers. 
TOMs are defined as technology-based objective measures 
provided by device-based clinical tests conducted in a standardized environment 
to have an objective assessment of specific behavior related to a 
movement disorder~\cite{espay2016technology}. 
TOMs can also include the tests self-administered by patients 
to monitor symptoms in everyday life. 
TOMs help the health professionals in assessing patient symptoms 
such that the quality of care can be improved~\cite{espay2016technology}.
The second input consists of preferences of the patients, who are the end users 
of the wearable devices, as shown in Figure~\ref{fig:health_scenario}b. 
The preferences include materials used in the device, form factor, and battery 
life. 
These inputs from HPs and patients ensure that the wearable devices developed 
meet the requirements of the users from the onset, rather than relying on customer feedback.

% Paragraph 3 goal: Development of the base platform by developers
The open source hardware and software is developed 
with the inputs from health professionals and users to address their 
requirements and needs.
The \textit{OpenHealth} platform also includes standard APIs, reference applications, and data sets. 
These APIs can be used by researchers to develop their own applications to detect and monitor TOMs without mastering the hardware design. 
An open source platform is important because it 
enables compatibility and standard comparisons of data. It can also enable 
the generation of common data sets for movement disorders, analogous to data sets 
such as the MNIST database\footnote{http://yann.lecun.com/exdb/mnist/} for 
image recognition.  
This can give a boost to the research in the area of movement disorders. 
In addition to the base open source platform, third parties can 
develop their own commercial applications as extensions to the base platform, 
as illustrated in Figure~\ref{fig:health_scenario}b.
%At the end of the development cycle, the open source wearable device is 
%provided to patients and HPs for use in home and clinical environments, respectively.

% Paragraph 4 goal: Usage of device by patients and HPs to supplement office 
% visits
The final step in the OpenHealth architecture is the real-world usage of 
wearable devices by patients and clinicians. The wearable device supplements 
the existing office visits, which could be weeks apart. In our 
\textit{OpenHealth} 
vision, the wearable device provides daily feedback of TOMs and other relevant 
parameters to both patients and their HPs. This allows HPs to 
monitor the symptoms of their patients in real-time, allowing them to make 
better 
therapeutic decisions. Similarly, patients can benefit by having access to 
daily feedback about their symptoms from both HPs and the algorithms on the 
wearable device. As a result of this daily feedback, large improvements to 
the quality of life of patients can be made.
%{\color{red}The device is sustainable and is ideal for prolonged continuous 
%use.} {\color{blue}Moreover, it is completely autonomous and needs minimum 
%human interaction for activity and health monitoring.}
%At the same time, there are 
%several challenges that need to be addressed for the success of this vision, 
%as 
%we describe in the next section.

%% file: files/opensource_hw.tex
\section{OpenHealth Release} \label{sec:hardware}

\begin{figure}[h]
	\centering
	\includegraphics[width=1.0\linewidth]{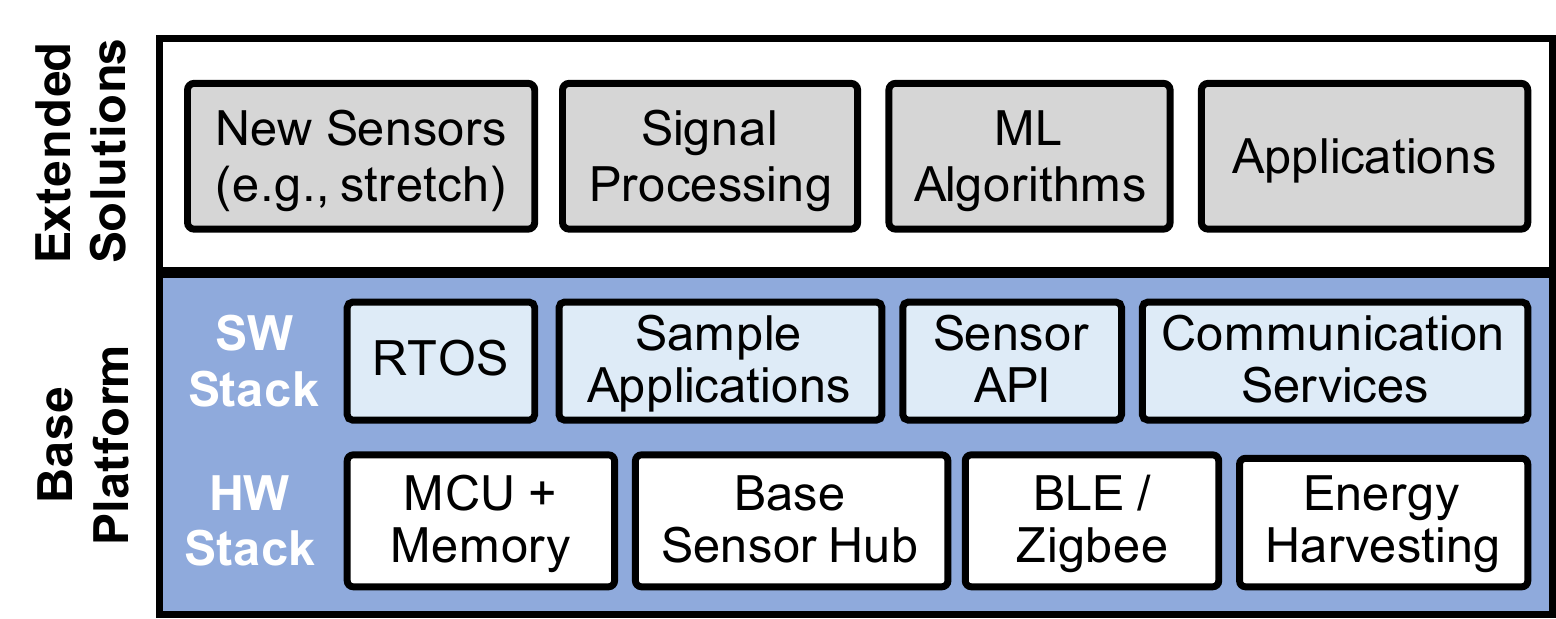}
	\caption{Overview of the wearable platform}
	\label{fig:wearable_platform}
\end{figure}

The main components of the \textit{OpenHealth} platform are shown in Figure~\ref{fig:wearable_platform}. The hardware 
stack in the base platform consists of most commonly used sensors, 
a microcontroller unit (MCU), radio and energy harvesting circuitry. 
Similarly, the base software stack consists of the real-time operating system~(RTOS), 
sensor APIs, communication services and reference applications. 
In addition to the base platform, the wearable platform 
can be extended with additional sensors, algorithms, and applications.

\subsection{The Base Hardware}\label{sec:base_hw}
\begin{figure}[t]
	\centering
	\includegraphics[width=0.9\linewidth]{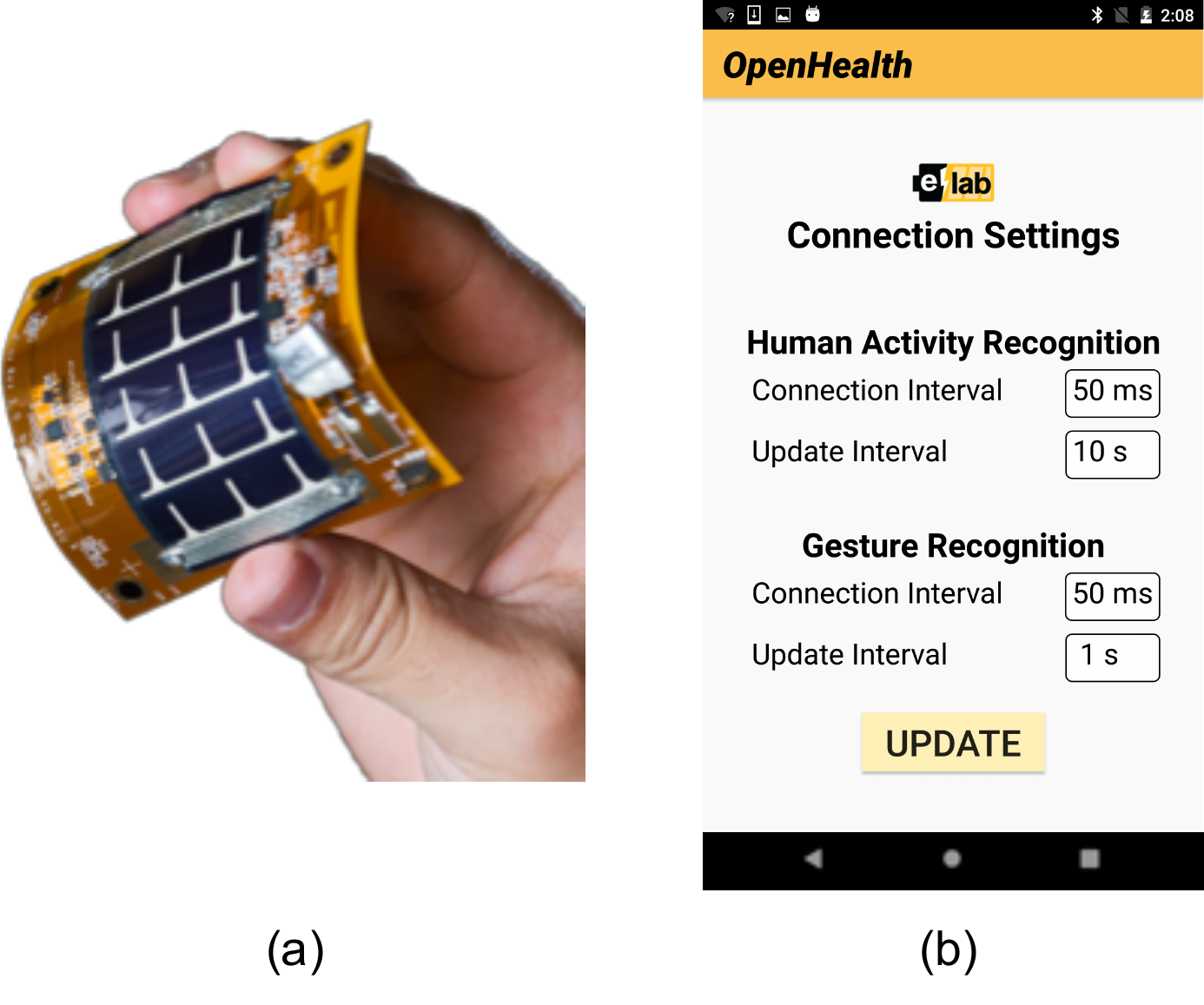}
	\caption{(a) Wearable device in flexible form factor (b) Screenshot of 
	the smartphone app used to control the device}
	\label{fig:prototype_photo}
\end{figure}

\noindent\textbf{Processing Unit:}
Texas Instruments~(TI) CC2650 
MCU\footnote{http://www.ti.com/product/CC2650} is the main processing unit in 
our base hardware. 
It consists of an ARM Cortex-M3 core with an operating frequency of 47~MHz. The 
MCU includes 20~KB of SRAM and 128~KB of programmable flash. In addition to the 
main Cortex-M3 core, it also includes a low power sensor controller that can 
run autonomously from the rest of the system. The sensor controller can be used 
to monitor sensors while the rest of the system is in a low power sleep state.
To ensure autonomous operation, the device is always on, but it waits in low power mode until it detects active motion. 
When motion is detected, the device wakes up and starts collecting sensor data. 
Then, the MCU executes the target application, 
such as human activity or gesture recognition, as detailed in Section~\ref{sec:applications}. 
The outputs are transmitted to a host device, such as a smartphone, 
without any user intervention after setting the connection settings with the host device, 
as illustrated in Figure~\ref{fig:prototype_photo}b. 
These settings control how often the wearable device synchronizes with the host. 
As mentioned before, on-board processing capability allows us to perform the processing of TOMs in real-time, thus 
eliminating the need to transfer the raw data.
%This helps in fulfilling the 
%local processing requirement.

\vspace{2mm}
\noindent\textbf{Sensor Unit:}
The sensor unit in our base wearable platform consists of the Invensense 
MPU-9250\footnote{https://www.invensense.com/products/motion-tracking/9-axis/mpu-9250/}
 motion sensor and electromyography~(EMG) sensors. The MPU 
consists of a three-axis accelerometer and a three-axis gyroscope. They are 
used to track the motion of the user wearing the device. Similarly, the EMG 
sensor is used to record the electrical activity produced by the muscles of the 
user. The sensors are connected to the MCU using an SPI interface, which makes 
it easy to interface additional sensors in future extensions.
We plan to add more sensors such as galvanic skin response and blood oxygen 
sensor in future versions of the device.
We note that adding new sensors may require changes in layout and power supply architecture on the device.

%Finally, the stretch sensor can be used to track the degree of bending of 
%knees 
%or elbow. The stretch sensor changes its capacitance as a function of the 
%strain applied to it. If a user wears the stretch sensor on the knees, it gets 
%stretched when the knee is bent, resulting in a change in the capacitance of 
%the sensor. This change can be used to detect the degree of bending of the 
%knee. This information can be used to evaluate the recovery of a patient after 
%a knee surgery. It can also be used to analyze the gait of the users.

\vspace{2mm}
\noindent\textbf{Communication:}
We use the integrated RF core in the 
TI CC2650 as the main communication module. It consists of a dedicated ARM 
Cortex-M0 to support communication tasks. The RF core can support the 
Bluetooth Low Energy~(BLE) and ZigBee protocols using a 2.4 GHz RF transceiver.
%In our applications, we use the BLE 
%protocol as the primary communication protocol. The wearable device is 
%connected to a host such as a smartphone or a computer. The BLE interface is 
%used to transmit important data for an application.
The BLE interface encrypts 
all the data that is sent over the air, thus ensuring the security and privacy 
of the user data. Moreover, BLE and Zigbee can be used to create a network of 
devices to make the platform scalable.

\vspace{1mm}
\noindent\textbf{Energy Harvesting Unit:}
We use energy harvesting as the primary source of energy in our wearable platform to enable sustainable operation. 
In particular, we use solar energy harvesting with the help of a PV-cell, as shown in Figure~\ref{fig:prototype_photo}. 
The PV-cell is connected to a maximum 
power point tracking charger that charges the Lithium-ion battery mounted on the device. 
The battery is used to store the harvested energy such that it can sustain autonomous operation when the harvested energy is below the energy requirement of the device~\cite{park2017flexible}.

% TI BQ25504

\vspace{1mm}
\noindent\textbf{Form factor:} The base hardware can be manufactured in both 
rigid and flexible forms. Figure~\ref{fig:prototype_photo}a shows the device in 
a flexible form factor. The flexible form factor is easy to wear as a patch on 
the body, which makes it comfortable to use
{for longer periods of time}.
We focus on incorporating a flexible form factor since 
	flexible electronics technology has the potential to change the landscape 
	in computing using stretchable and bendable platforms. 
	Flexible hybrid electronics~(FHE)~\cite{gupta2017flexibility} can enable powerful computing abilities in a stretchable and bendable form 
	factor by taking advantage of the computing abilities of rigid processors. 
	As a result, FHE technology allows us to perform local processing of TOMs, 
	while maintaining a form factor that is comfortable and easy to wear as a patch.
The rigid form factor is useful when a more rugged operation is 
required, such as mounting the device on a shoe. Both the form factors use 
PV-cells to harvest energy from the ambient light, fulfilling the energy 
harvesting requirement.

%\textcolor{red}{Refer to Ujjwal's paper. Also, I pasted a paragraph from a 
%proposal. 
%We can use a shorter version here: }
%\textcolor{blue}{
%Flexible electronics technology has the potential to transform computing by enabling bendable and stretchable wearable systems at a low cost [149]. 
%Physical flexibility combined with cost and weight advantages opens a wide range of form factors and application areas, including wearable electronics, medical sensing, rollable displays, and Internet of Things (IoT) [97, 105, 127, 130]. However, the performance and capabilities of purely
%flexible electronics are currently very limited [101, 134]. For example, silicon technology offers 14nm feature size with over 1GHz frequency, whereas the feature sizes of thin-film transistors (TFT) range from 8 $\mu$m to 50 $\mu$m [62], and frequencies hardly reach 10MHz [74]. Emerging flexible hybrid electronics (FHE) technology addresses this problem by integrating rigid Silicon chips and printed electronics [19, 128]. FHE can be used to coalesce rigid and flexible resources judiciously to bridge the gap between today’s complex systems and the advantages of flexible electronics. Hence, FHE can drive the next big leap forward in the form factor design, similar to the shift from desktop and laptop computers to hand-held devices.}
%

%% file: files/software_stack.tex
\subsection{The Base Firmware} \label{sec:software_stack}
\noindent\textbf{Operating System:}
The primary operating system consists of a thread-based, real-time operating 
system~(RTOS) from Texus Instruments. The RTOS is responsible for scheduling and maintaining 
all the software tasks running in the system. The RTOS also provides drivers 
such as I2C, SPI, and UART to interface with the peripherals and sensors 
connected to the MCU. 
We employ the SPI interface to 
connect the motion sensors to the MCU.

\vspace{1mm}
\noindent\textbf{Sensor API:}
The sensor API functions as the intermediate interface between low-level 
drivers and the application. In our software architecture, the 
sensor API provides functionality to control the sensors using the I2C 
or SPI interfaces. Specifically, it provides standard functions that allow the 
application to control the registers of the sensors and read data from each 
sensor. The modular nature of the sensor API interface allows the developers to 
easily add new sensors.

\vspace{1mm}
\noindent\textbf{Communication services:}
The communication services consist of the BLE and Zigbee protocol 
stacks. The protocol stacks run on the RF core in the MCU, independently of the 
application. Whenever the application has to send data to another device, it 
sends a message to the stack which transmits the message over the air.
All the data transferred using BLE and ZigBee is encrypted to ensure the 
security of the data.

%\textcolor{red}{In addition to these components, the firmware contains application routines 
% to generate the TOMs approved by HPs. In the next section, we describe some example applications using the proposed wearable platform.}
%
\vspace{1mm}
\noindent\textbf{Ease of development:}
We use the TI CC2650 MCU to ensure that tools needed for development of 
software for the wearable platform are easily accessible. Hence, \textit{OpenHealth} users can use all the software tools available for the TI MCU. This includes software development kits, debugging 
kits, and RTOS libraries. In addition to the software tools provided by TI, we 
provide standard APIs to use the sensors in the \textit{OpenHealth} platform.
Finally, we plan to create a forum that will enable the users of the platform
to discuss new applications, designs, and solutions to potential issues.

\subsection{Public Release}
The design files and firmware for the base platform are available for download 
at the \textit{OpenHealth} web page\footnote{https://sites.google.com/view/openhealth-wearable-health/home}. 
%Alternatively, users can extend the design to add additional components as per their application requirements. 
In addition to HW design files, the public release also contains the base firmware that includes the RTOS, sensor API, communication libraries. 
To facilitate development effort and new research, we also include the reference applications presented in Section~\ref{sec:applications}, 
and anonymous data sets collected for these applications.
Users of the wearable platform will also be able to upload new 
data sets, thus helping in the creation of a common data repository for movement disorders. 
We plan to maintain an \textit{OpenHealth} Wiki-page that will describe the steps 
required for manufacturing the base platform and reference applications, 
as well as answering frequently asked questions. 

%% file: files/applications.tex
\section{Example Application Domains} \label{sec:applications}
%This sections first presents example applications that are released along with 
%the open source wearable platform.

%\noindent\textbf{Human Activity Recognition:}
The \textit{OpenHealth} platform can be used to implement a variety of wearable applications 
ranging from fitness tracking to continuous health monitoring of patients with movement disorders.
One of our focus application areas is diagnosis and monitoring of patients with PD. 
For example, body motion analysis, response to therapy and motor fluctuation 
monitoring of PD can be achieved with 3-axis accelerometers and gyroscopes available in the base platform. 
This section illustrates two reference applications included with the \textit{OpenHealth} release. 
Adding more sensors like sweat sensors, heart rate sensors and EEG can enable monitoring of nonmotor symptoms and progression~\cite{espay2016technology}.
A more detailed analysis of the data from wearable sensors and 
development of new algorithms for human activity recognition is left as future work.

\subsection{Human Activity Recognition (HAR)}
\begin{figure}[ht]
	\centering
	\includegraphics[width=0.95\linewidth]{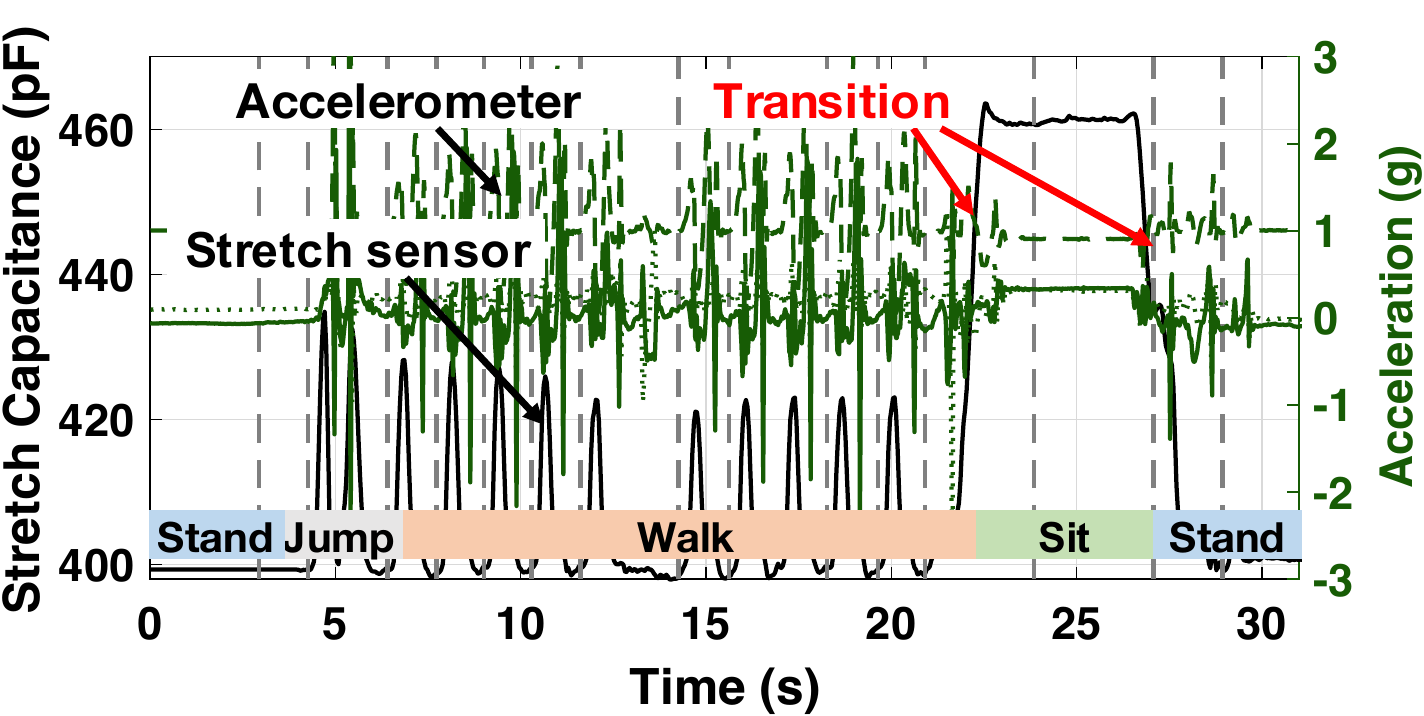}
	\caption{Sequence of activities in the HAR application}
	\label{fig:har}
\end{figure}

One of the first steps in the treatment of movement disorders is to understand what the patients are doing~\cite{daneault2018could}.
This objective is achieved through HAR algorithms, 
which aim to identify the user activity, such as, standing, walking, and jogging, by processing sensor data. 
Since efficient HAR implementations can provide valuable insights to both health professionals and patients, 
we include it as one of the two reference applications. 
To implement our HAR application,  we employ the base platform, 
which contains the MPU-9250 motion sensor, along with a wearable stretch 
sensor~\cite{bhat2018online}. We use a combination of accelerometer 
and stretch 
sensors since it provides 10\% higher recognition accuracy than using either 
sensor alone.
The 3-axis accelerometer in the motion sensor is placed near the ankle 
to capture the swing of one of the legs. 
The stretch sensor is used on a knee sleeve to capture the degree of bending of the knee. 
During this work, we performed 58 unique experiments with 9 users 
and collected data for the activities 
summarized in Table~\ref{tab:accuracy_har}. 
Then, the data is used to train a programmable neural network 
that can recognize these activities in real-time. 
A sample snapshot that contains the data for \textit{Stand, Jump, Walk,} and \textit{Sit} 
activities is shown in Figure~\ref{fig:har}. 
Our experimental evaluation shows that we achieve an accuracy 
greater than 90\% for all the activities, as shown in 
Table~\ref{tab:accuracy_har}.
The application consumes about 12.5 mW power for recognizing a single activity. 
\textit{The user data as well as the implementation of the HAR application 
will be included in the \textit{OpenHealth} release}.

\begin{table}[ht]
	\centering
	\caption{Accuracy of the HAR application}
	\label{tab:accuracy_har}
	\begin{tabular}{@{}lcr@{}}
		\toprule
		Activity    & \# Correct / \# Total Segments &Accuracy (\%) \\ \midrule
		Drive       & 154 / 155 &99.4          \\ 
		Jump        & 169 / 181 &93.4          \\
		Lie Down    & 204 / 204 &100           \\
		Sit         & 385 / 394 &97.7          \\
		Stand       & 345 / 350 &98.6          \\
		Walk        & 794 / 806 &98.5          \\
		Transitions & 115 / 127 &90.5          \\ \bottomrule
	\end{tabular}
\end{table}

%
%

%\noindent\textbf{Gesture Recognition:}

\subsection{Gesture Recognition}
Gesture recognition is another application that is very useful 
in the context of health monitoring. 
Gesture recognition can be used in applications 
such as gesture-based control and interaction with assistive devices. 
We implemented a gesture recognition application 
that uses the  accelerometer sensor in the base platform.
The device is mounted on the wrist of the 
user to record the accelerometer data when the user is performing a gesture. 
Using a NN classifier, our application recognizes gestures, 
such as \textit{up, down, left} and \textit{right},  
which can be  used to control assistive devices. 
Experimental evaluations using seven users show 
that the wearable device can recognize gestures with an accuracy of 98.6\% 
while having an active power consumption of about 10 mW.
Our implementation of the gesture recognition 
application~\cite{park2018optimizing} and the corresponding data set is available 
with the release of the wearable device.

%% file: files/conclusion_future_research.tex
\section{Conclusion and Future Directions}\label{sec:conclusion}

This paper presented the \textit{OpenHealth} platform for open source health monitoring. 
It discussed the need for wearable health 
monitoring and the challenges faced by wearable devices before their widespread 
adoption. Then, it presented the hardware and software details of the 
\textit{OpenHealth} platform. 
Finally, we provided example applications for 
human activity and gesture recognition using the proposed wearable platform.

In order to assess whether the wearable device is able to address the 
challenges and meet our vision, we will conduct extensive user studies with movement 
disorder patients.
We also plan to design custom SoC with reconfigurable NN accelerators to reduce the 
power footprint into $\mu$W range. Furthermore, we will integrate additional 
modalities of energy harvesting such as body heat and body motion. These will 
be key enablers to sustainable and maintenance free operation. 

%% file: opensource_hw_date.bbl
% Generated by IEEEtranS.bst, version: 1.14 (2015/08/26)